\documentclass[submission,copyright,creativecommons]{eptcs}
\providecommand{\event}{FMAS'22} 
\usepackage[utf8]{inputenc}
\usepackage{graphicx}
\usepackage{lipsum}

\usepackage{comment} 

\usepackage{cite}
\usepackage{xcolor}
\usepackage{hyperref}
\usepackage{amsmath,amssymb,amsfonts}
\usepackage{wrapfig}
\usepackage{float}
\usepackage{enumitem}
\usepackage{tikz}
\usetikzlibrary{arrows,shapes,automata,petri,backgrounds,calc,intersections,decorations.pathreplacing}

\newcommand{\exmodel}{explanation model}
\newcommand{\sysmodel}{system model}
\newcommand{\sys}{\textsf{SysModel}}
\newcommand{\expath}{explanation path}
\newcommand{\modelone}{\textsf{EM1}}
\newcommand{\modeltwo}{\textsf{EM2}}
\newcommand{\modelthree}{\textsf{EM3}}
\newcommand{\modelthreex}{\textsf{EM3(X)}}
\newcommand{\modelthreey}{\textsf{EM3(Y)}}

\newcommand{\modelfour}{\textsf{EM4}}
\newcommand{\modelfourx}{\textsf{EM4(X)}}

\newcommand{\modelfivex}{\textsf{EM5(X)}}

\newcommand{\modelfivexone}{\textsf{EM5(x$_1$)}}
\newcommand{\modelfivextwo}{\textsf{EM5(x$_2$)}}
\newcommand{\modelfivexN}{\textsf{EM5(x$_n$)}}

\newcommand{\ego}{\ensuremath{\mathit{ego}}}

\title{From Specification Models to Explanation Models: An Extraction and Refinement Process for Timed Automata}
\def\titlerunning{From Specification Models to Explanation Models}

\author{
Maike Schwammberger\thanks{M.\ Schwammberger was supported by the German Research Council (DFG) in the PIRE Projects SD-SSCPS and ISCE-ACPS under grant no.\ FR 2715/4-1 and FR 2715/5-1.} 
\institute{University of Oldenburg,\\Oldenburg, Germany}
\email{schwammberger@informatik.uni-oldenburg.de}
\and
Verena Klös 
\institute{TU Berlin,\\Berlin, Germany}
\email{verena.kloes@tu-berlin.de}
}
\def\authorrunning{Maike Schwammberger and Verena Klös}

\begin{document}

\maketitle

\begin{abstract}
    Autonomous systems control many tasks in our daily lives. To increase trust in those systems and safety of the interaction between humans and autonomous systems, the system behaviour and reasons for autonomous decision should be explained to users, experts and public authorities. One way to provide such explanations is to use behavioural models to generate context- and user-specific explanations at run-time. However, this comes at the cost of higher modelling effort as additional models need to be constructed. In this paper, we propose a high-level process to extract such \textit{explanation models} from system models, and to subsequently refine these towards specific users, explanation purposes and situations. By this, we enable the reuse of specification models for integrating self-explanation capabilities into systems. We showcase our approach using a running example from the autonomous driving domain.
\end{abstract}

\section{Introduction}\label{sec:introduction}

Nowadays, autonomous systems control many areas of our daily lives, e.g. tasks in transportation, medicine or industry. These tasks require context-awareness, and errors and failures might have severe consequences. 
To increase trust in and safety of those systems, the system behaviour and reasons for autonomous decision should be explained to users, experts and public authorities.
Explainability has become a hot research topic in the area of artificial intelligence~\cite{goebel2018explainable, Holzinger2022}, but also becomes more important for any autonomous system \cite{anjomshoae2019explainable, greenyer2019explainable, Ziesche21, weyns2021research}. 

One way to explain the behaviour of autonomous systems is to directly encode the generation of explanations into the system code (e.g., done in \cite{agrawal2021explaining}). However, this approach only allows for generating explanations for behaviours that have been classified as relevant at design time. A more flexible approach, that also facilitates the adjustment of explanations to the explanation recipient and context, is the generation of explanations at run-time. 
While the first approach can profit from direct access to the encoded system decisions, the latter needs models of the system to identify the events and decisions that led to the current system behaviour. We call these models \emph{\exmodel s}. These models are either manually constructed at design time (e.g., in \cite{Getal18}), or learned from data (e.g., in \cite{Ziesche21, plambeck2022}). In \cite{Betal19}, we have proposed the MAB-EX framework that uses \exmodel s to generate explanations on demand, at run-time. There, we have defined explanation models as follows:
\begin{quote}
    \emph{An explanation model is a behavioural model of the system that captures causal relationships between events and system reactions. It allows for identifying possible causes for the behaviour that needs to be explained, e.g., traces
of events that may lead to the behaviour. It may also allow
for look-ahead simulation to enable answering questions like
“What happens if ... ?” or “When will ... be possible again?”.}~\cite[p. 544-545]{Betal19}
\end{quote}

The flexibility of model-based explanations comes at the cost of higher modelling effort as \exmodel s are additional models that need to be constructed. 
These models have to capture causal relationships between events and system reactions. They should enable the identification of the current state from monitored data and back-tracing with monitored data to identify the chain of causes. An optional requirement is to allow for simulation to answer questions on alternatives and possible future behaviour. Thus to construct them, deep insights into the system behaviour in different situations and contexts is needed.
Building them is a new challenge, that has not yet been researched thoroughly. 

\textit{Contribution.}
In this paper, we propose to reduce the modelling effort by 
extracting \exmodel s from formal system models that were used to specify and verify the system behaviour at design time and to subsequently refine these for users, explanation purposes and situations. The notion of ``refinement'' that we use within this paper differs from the well-known refinement concept from formal methods. Instead, with the term refinement, we describe a process where a coarse \exmodel{} is adapted to a more detailed structure. Thus with our meaning of refinement, the explanation-capability of an \exmodel{} is increased.
For now, we focus on an approach to extract initial models from timed automata \cite{AD94} that is based on the vision paper \cite{Sch21-quest}. Furthermore, we propose a high-level extraction and refinement process that describes which information should be hidden or added for which purpose. We also sketch continuous refinement at run-time to allow for user-specific adjustments and for integrating new information. We illustrate the steps of our high-level approach with a running example from the domain of autonomous driving. Note that the aim of this paper is to present the idea and general process of extracting explanation models from system models but that we do not yet give an implementation.

The main advantages of our high-level extraction and refinement process are:
\begin{itemize}
   \item Model reuse: By extracting an \exmodel{} from an existing \sysmodel, we enable the reuse of design-time models
   \item Formal Foundation: By using a formal, verifiably correct system model like a timed automaton, we avoid generating explanations from ambiguous natural language (e.g. directly from requirements).
    \item Modularity: With the help of the created \exmodel{} and our MAB-EX framework, we can introduce self-explainability to existing systems and also facilitate system updates at run-time, as the explanation model can also be updated at run-time.
    \item User-specificity: The output of our framework is an \exmodel{} that takes different types of explainees into account. This is reasonable as, for instance, an engineer might need differently detailed explanations than an end-user.
\end{itemize}

\textit{Outline.}
In the following, we briefly present the MAB-EX framework~\cite{Betal19} for self-explaining systems that uses explanation models to construct explanations on demand. Afterwards, we present our explanation model extraction and refinement process in Sect.\ref{sec:creation-process} and introduce our running example in Sect.~\ref{sec:case-study}. We explain the extraction phase in detail in Sect.~\ref{sec:model-extraction} and the refinement phase in Sect.~\ref{sec:model-refinement}. In Sect.~\ref{sec:updates}, we discuss reasons for run-time adaptation of \exmodel s. We discuss our approach in Sect.~\ref{sec:evaluation} and conclude the paper in Sect.~\ref{sec:conclusion}.

\section{Preliminaries: MAB-EX Framework for Self-explaining Systems}\label{sec:mabex}
\begin{figure}[htbp]
    \centering
    \includegraphics[width=5.5cm]{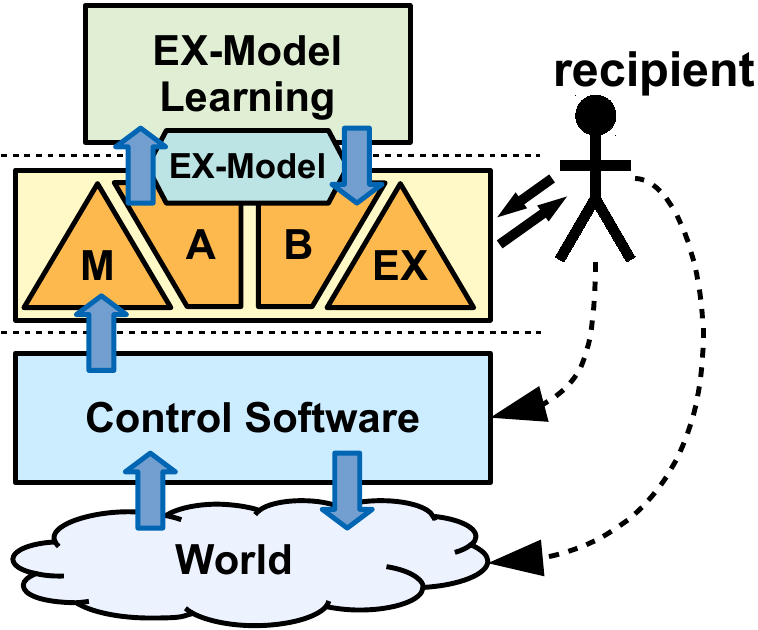}
    \caption{MAB-EX Loop from \cite{Betal19}.}
    \label{fig:mabex}
\end{figure}

To enable the design of self-explainable systems, we have proposed the MAB-EX Loop (Monitor, Analyse, Build, Explain) in previous work \cite{Betal19}. The main idea of this reference framework is to adopt the main principles of the well-known MAPE Loop~\cite{ibm2005architectural} for self-adaptive systems to build self-explaining systems. We depict the MAB-EX Loop in Fig.~\ref{fig:mabex}.

The MAB-EX Loop monitors and analyses the behaviour of a system and decides whether the user (or another stakeholder) requires an explanation. This decision can depend on different factors: on the current context, on the user's experience with similar situations, or on the characteristics of the situation (e.g., rareness). For more factors that influence the need for an explanation, we refer to \cite{Sadeghi21}. If such an explanation need was detected, the next phases of the loop then build an explanation from explanation models and convey this explanation in a suitable way to the stakeholder. Building an explanation and actually presenting the explanation are separated in the MAB-EX Loop to allow for individual explanations for different stakeholders and contexts.


Possible implementations for an explanation model, that were discussed and illustrated in the paper, are (fault/decision) trees that connect observations to possible reasons and executable behaviour models, 
e.g., state machines.
In the \emph{Build}-phase, the observed behaviour is used to identify the current state in the explanation model and the events that have led to this state. These events form a trace, which we refer to as \emph{explanation path} in the remainder of the paper. This path is an internal representation of the explanation that is further processed in the \emph{EX}-phase of the loop, e.g. enriched with further information and transformed into a suitable presentation format that is given to the \emph{explainee} (i.e. the recipient of the explanation).

Note that dynamically building explanations makes most sense for complex context-sensitive, maybe self-adaptive or learning systems where it is impossible to predict all situations that require an explanation beforehand. Furthermore, by adding an explanation layer that follows the MAB-EX framework, it is possible to add explanation capabilities to an existing system.



In this paper, we focus on constructing explanation models that can be used in the MAB-EX framework. Additionally, and in contrast to \cite{Betal19}, we propose to already tailor the explanation models to different types of explainee. By this, the generated explanation path is reduced to events that are relevant for the explainee and the subsequent processing in the \emph{EX}-phase can be simplified.

\section{Explanation Model Extraction and Refinement Process}\label{sec:creation-process}
Our overall goal is to introduce a multi-level extraction and refinement process for making a variety of existing systems self-explainable. For this, we suggest that an \emph{\exmodel} is extracted from an existing \emph{\sysmodel} and then further refined to cope with different explanation purposes and explainees. As motivated in Sect.~\ref{sec:mabex}, an \exmodel{} is a causal structure that connects system actions with their reasons (i.e. events preceding the action). 
From such an \exmodel, explanations may be generated on demand, at run-time. 
In the following, we motivate some key concepts of our approach, before we introduce the actual \exmodel{} extraction and refinement process.


\subsection{Type of \sysmodel. } 
In our terminology, a \sysmodel{} is some technical description of the system's behaviour, goals and general functionality.
We envision that our approach is not limited to a specific type of system: it summarises engineering steps needed to derive an \exmodel{} from a \sysmodel. The extraction of an initial \exmodel{} is done by exploiting the syntactical and semantical structure of the \sysmodel. In this paper, we use timed automata system models. Similar steps are also applicable for other types of system models. However, the
notion of causality in different models might be very different, and influences the steps in our process.


\subsection{Types of explanations. } Across domains and throughout research disciplines, different types of explanations are examined for their applicability and usefulness. In the context of semi-autonomous driving, the authors of \cite{Ketal15} discovered that ``why'' (e.g., ``Obstacle ahead'') explanations are preferred over ``how'' explanations (e.g., ``The car is braking'') by drivers and led to a better driving performance. A combination of both ``why'' and ``how'' explanations led to the safest driving performance. Research results of \cite{LDA09} substantiate these findings as the authors describe that 
``why'' explanations can improve a user's trust in a system and are 
more easily understood by the user than ``why not'' explanations. 

We follow these approaches and consider reasoning traces that combine ``how'' and ``why'' explanation types, e.g.: ``The car did brake (``how''), because an obstacle is ahead (``why''). In our case, these reasoning traces relate to the explanation paths, that we introduced in Sect.~\ref{sec:mabex}.

\subsection{Explainee types and explanation purposes. }
A key feature of our approach is that we take different \emph{types of explainees} into account. The term explainee comprises the recipient of an explanation. In \cite{Ketal19}, the authors also suggest, as a first requirement for explainability, to characterise traits of different types of explainees. Such different types of explainees can, for example, be deduced from the stakeholders that are identified in the requirements engineering phase of the system engineering process \cite{reqeng17}.
Here, we postulate that differently detailed explanation models are needed for different explainee types and for different \emph{explanation purposes}. By explanation purpose, we mean the topic or circumstance that needs to be explained. Examples of explainee types and explanation purposes are:
\begin{itemize}\label{note:explainee-types}
    \item an end-user, who needs to cooperate with the system or who simply wishes to understand some system behaviour;
    \item an engineer, who needs to understand a system failure; 
    \item a lawyer or the general public, who need to figure out whether a system is responsible for an accident; or
    \item another system that interacts with the considered system, e.g. two autonomous cars from different manufacturers that need to cooperate.
\end{itemize}
Note that mechanisms for exploring the needs of different explainee types are out of the scope of this paper, but we refer to \cite{Ketal19} for this, where requirements for explainability in general are explored.


\subsection{The multi-level extraction and refinement process -- Overview. } 
\begin{figure}[htbp]
    \includegraphics[width=\linewidth, trim=0 0.1cm 0 0, clip]{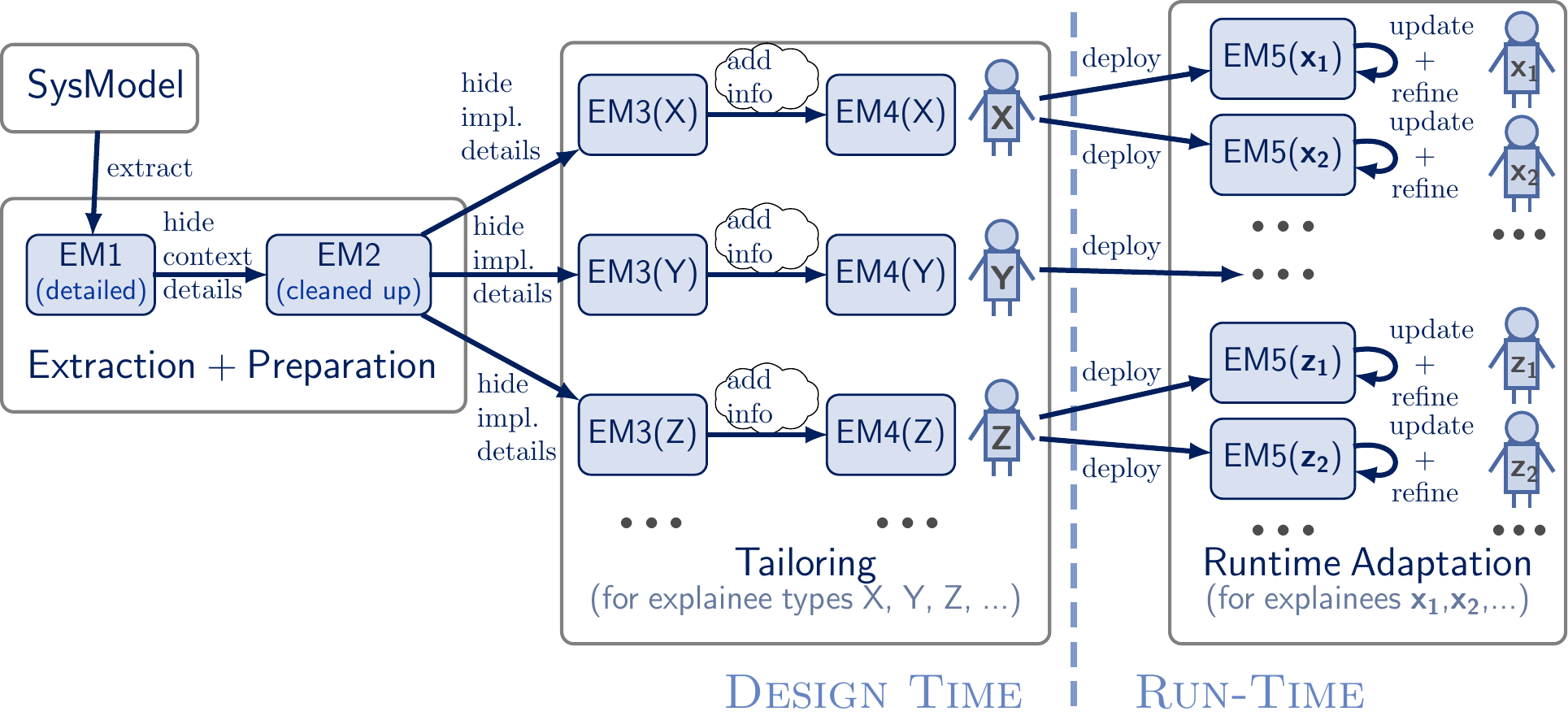}
    \caption{Overview of the explanation model creation process.}
    \label{fig:ex-models}
\end{figure}

We give an overview of our framework in Fig.~\ref{fig:ex-models}. We distinguish three different phases (each having a separate box in Fig.~\ref{fig:ex-models}) which lead from a formal \sysmodel{} \sys{} to a user-specific \exmodel{} \modelfivexone, \modelfivextwo, ..., \modelfivexN, for specific individual explainees \textsf{x$_1$}, \textsf{x$_2$}, ..., \textsf{x$_n$}.
We list and briefly describe these three phases in the following and give more details in Sects.~\ref{sec:model-extraction} to \ref{sec:updates}, respectively.

Note that each of the steps that we propose is meant to optimise an initially extracted \exmodel. This means that each of the intermediate \exmodel s is already functional on its own. Such intermediate models might be useful for an integration into other approaches and implementations of explainability. If necessary, for example to achieve a fully automatic extraction and refinement, it is possible to skip some of the steps, although then the resulting \exmodel{} may not be optimised towards giving explanations. 

\textbf{Phase 1: Extraction and Preparation (Sect.~\ref{sec:model-extraction}).} This phase of the framework contains two different steps: Extraction of a first version of the \exmodel{} (\modelone{}) from the \sysmodel{} and a preparation of this extracted model for further phases.

\textit{Step 1: Extraction.}
In the case of a \sysmodel{} that is modelled as an automaton, this could, e.g., be done by connecting actions with their causes, as is done in the vision paper \cite{Sch21-quest}.
However, \cite{Sch21-quest} does not give any methodology for retrieving the explanation model from the crossing controller and omits details in the model without giving reasons for that. The data contained in such an 
extracted \exmodel{} \modelone{} is not yet filtered and contains reasons for all details from the \sysmodel, possibly also details that are irrelevant for the explanation topic.

\textit{Step 2: Preparation.}\label{page:preparation-step} As an unnecessarily large \exmodel{} is not desirable (e.g. because of too large computation times for time-critical explanations), some details are hidden in the preparation step of this phase, leading to \modeltwo. Let us consider an example from a smart factory, where an autonomous robot fetches and delivers parcels between different work stations: We assume that the behaviour of the robot is modelled via two different \sysmodel s that are working in parallel:
\begin{itemize}
    \item System model 1: Identify, grab or lay down a parcel
    \item System model 2: Move, stop, turn left, turn right
\end{itemize}
If an engineer is interested in explanations for the acceleration or deceleration of the robot (i.e. reasons for the actions ``move'' and ``stop''), it might be not necessary to keep the reasons for turning left and right within the \exmodel{} \modelone. Thus, these parts from \modelone{} are hidden upon the creation of \modeltwo{} in this example.

\textbf{Phase 2: Tailoring (Sect.~\ref{sec:model-refinement}).} The second phase is again parted into two steps: Firstly, explainee specific parts of \modeltwo{} are hidden, and secondly the resulting model \modelthree{} is enriched by details that were missing in the \sysmodel. These details are added by human experts or could be extracted from requirements specifications or other (formalised) documents.

\textit{Step 1: Hiding explainee specific details.} For different explainees, differently detailed information is needed for an explanation and hence also for the \exmodel. For instance, while exact values of internal data variables might be of interest for an engineer explainee type \textsf{X}, e.g. for analysing the system, these values will have little to no meaning for an end-user explainee type \textsf{Y} (\textsf{X},\textsf{Y} $\in$ \textsf{ExplaineeTypes}).
Thus, in this step, we start with \modeltwo{} and hide those details, that are irrelevant for a specific type of explainee \textsf{X} or \textsf{Y}. The result of this first step are \exmodel s \modelthreex and \modelthreey, for respective explainee types \textsf{X}, \textsf{Y}.

\textit{Step 2: Add expert information.} Specific information, e.g., that an autonomous vehicle is breaking because of a traffic rule, might not be directly included in \sys. Instead, it might be encoded that the car brakes because the value of a variable $b$ is not allowed to exceed a constant value $c$. Thus, we enrich abstract internal system data with environmental information. Such information may be, e.g., gathered from the system requirements or a system engineer. Although this step requires some high-level insights into reasons for the system behaviour, it is still less complex than having to create the whole explanation model from scratch.
The needed level of detail of this information is again dependant on the explainee, as, e.g., the value of a variable might have a sufficient meaning for an engineer or another system, but not for an end-user. The result of phase 2 is explanation model \modelfourx. 
Note that, when involving a human expert, this step can only be semi-automated. However, there exist possibilities for automating this step: additional information may be extracted from formalised requirements (cf.~\cite{FLSM22}), formalised traffic rule books (cf.~\cite{CDF22}) and similar sources.

\textbf{Phase 3: Run-time Adaptation (Sect.~\ref{sec:updates}).}
In our framework, the result of phase 2, the \exmodel{} \modelfourx, is meant to be deployed in a system to make it self-explainable for explainee type \textsf{X}. During run-time, 
further updates and adaptations can be necessary to consider evolving personal preferences of individual explainees or changes in the system behaviour due to self-adaptation, learning or software updates. The initially deployed \exmodel{} \modelfivexone{} for explainee \textsf{x$_1$} $\in$ \textsf{X} in this phase is the same as \modelfourx. 

Note that, depending on the used \sysmodel{} \sys, the explanation purpose and the considered type of explainee, it may be that some of the steps are omitted. For instance, if all of the details in \modeltwo{} are needed for an engineer explainee type, we may not hide details and \modeltwo is also used as \modelthree.

\section{Set-Up for Running Example}\label{sec:case-study}
We showcase the use and key features of our framework by using a specific running example with a specific \sysmodel{} \sys{} in this paper: an extended timed automaton. Our example is from the domain of autonomous driving and considers a crossing protocol for autonomous turn manoeuvres at urban intersections. It is a simplified adaptation of a crossing protocol that was introduced in \cite{Sch18-TCS,BS19}. 
We chose this example, as it comprises a concise, formal, definition of traffic manoeuvres and its soundness and the key features of the protocol have already been formally proven (e.g. safety). Furthermore, timed automata \cite{AD94} are a popular mechanism for the specification of various system types. With this, our running example also conforms to other systems.
We decided to simplify the protocol, as the introduction to the underlying formalism \emph{Automotive Controlling Timed Automata (ACTA)} with its traffic logic \emph{Urban Multi-lane Spatial Logic (UMLSL)} from \cite{Sch18-TCS, BS19} would be beyond the scope of this paper, and the formal details are not necessary for the purpose of the running example: Showcasing and illustrating our extraction and refinement process from Sect.~\ref{sec:creation-process}.
We refer to \cite{BS19, Sch18-TCS} for the formal details.

\begin{figure}[ht]
    \centering
    \includegraphics[width=8.5cm, trim=1.4cm 1.75cm 1.1cm 1.55cm, clip]{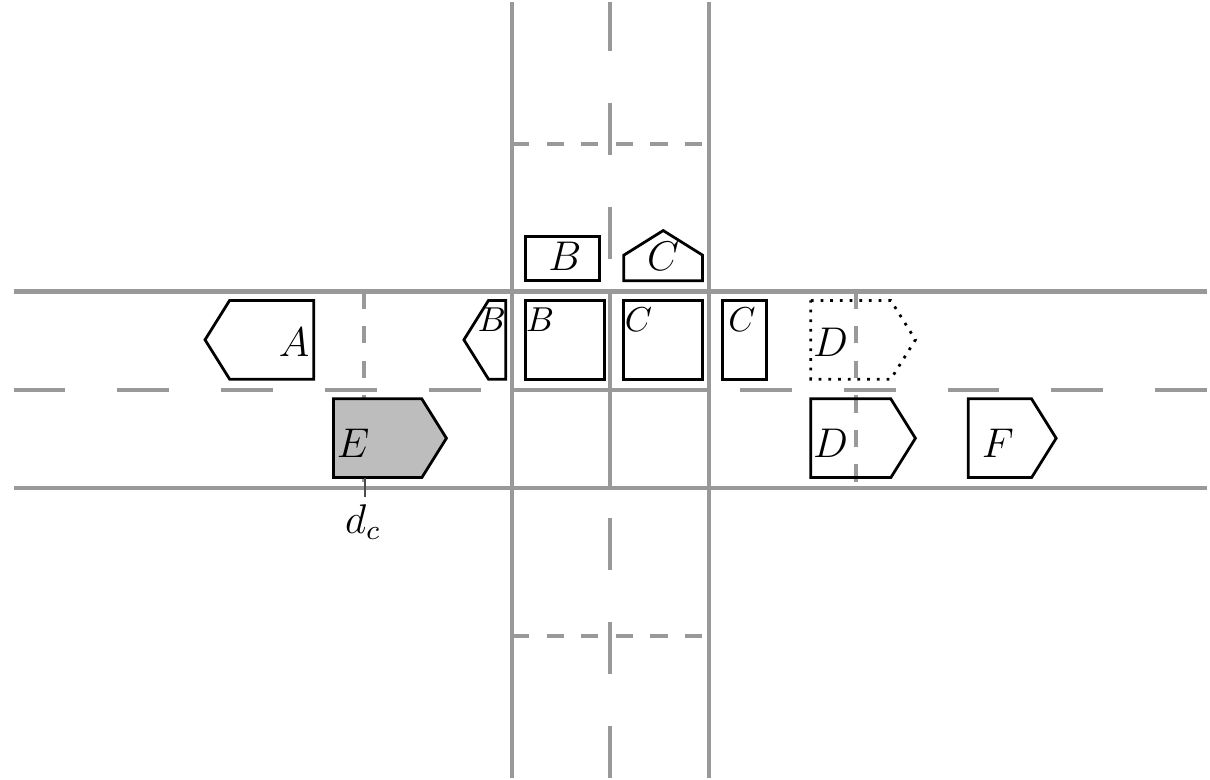}
    \vspace{-0.0cm}
    \caption{Example traffic situation where cars $A$ to $F$ meet at an urban intersection.}
    \label{fig:pre:intersection}
\end{figure}

A first visionary approach for extracting an explanation model for the crossing controller from \cite{BS19} was introduced in \cite{Sch21-quest}. 
An example for a traffic situation that can be handled by the protocol is depicted in Fig.~\ref{fig:pre:intersection}, where an \ego{} car $E$ approaches an intersection at which cars $B$ and $C$ are currently performing turn manoeuvres. With \emph{\ego{} car}, we refer to the car from whose viewpoint the traffic situation is considered.
The \ego{} car's goal is to turn left at the intersection.

\begin{figure}[htbp]
\vspace{-0.0cm}
    \centering
    \includegraphics[width=0.75\linewidth]{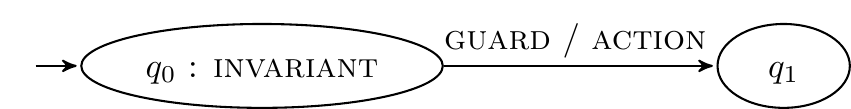}
    \vspace{-0.cm}
    \caption{Schematic timed automaton transition.}
    \label{fig:case-study:transition}
    \vspace{-0.1cm}
\end{figure}

The simplified crossing controller that we present in Sect.~\ref{sec:crossing-controller} is an extended timed automaton. For extracting an explanation model from a timed automaton, we suggest to identify and connect \emph{actions} and \emph{reasons} in the automaton. For this, consider the schematic timed automaton transition that we depict in Fig.~\ref{fig:case-study:transition}. Actions are elements that appear after $/$ and their reasons are the transition guards before $/$, as well as possible invariants in the starting location of the transition. Actions can be communicating events or operations on data and clock variables, which can also be specified in method code in some dialects of timed automata. Guards specify logical propositions on data and clock variables, which can also be encapsulated in method code, or specify communication events that have to be received. Invariants describe data and clock constraints that must not be violated while staying in the location.


\subsection{Crossing Controller Protocol}\label{sec:crossing-controller}
\begin{figure}[htbp]
    \centering
    \includegraphics[width=\linewidth, trim=1.3cm 0 0 1.3cm, clip]{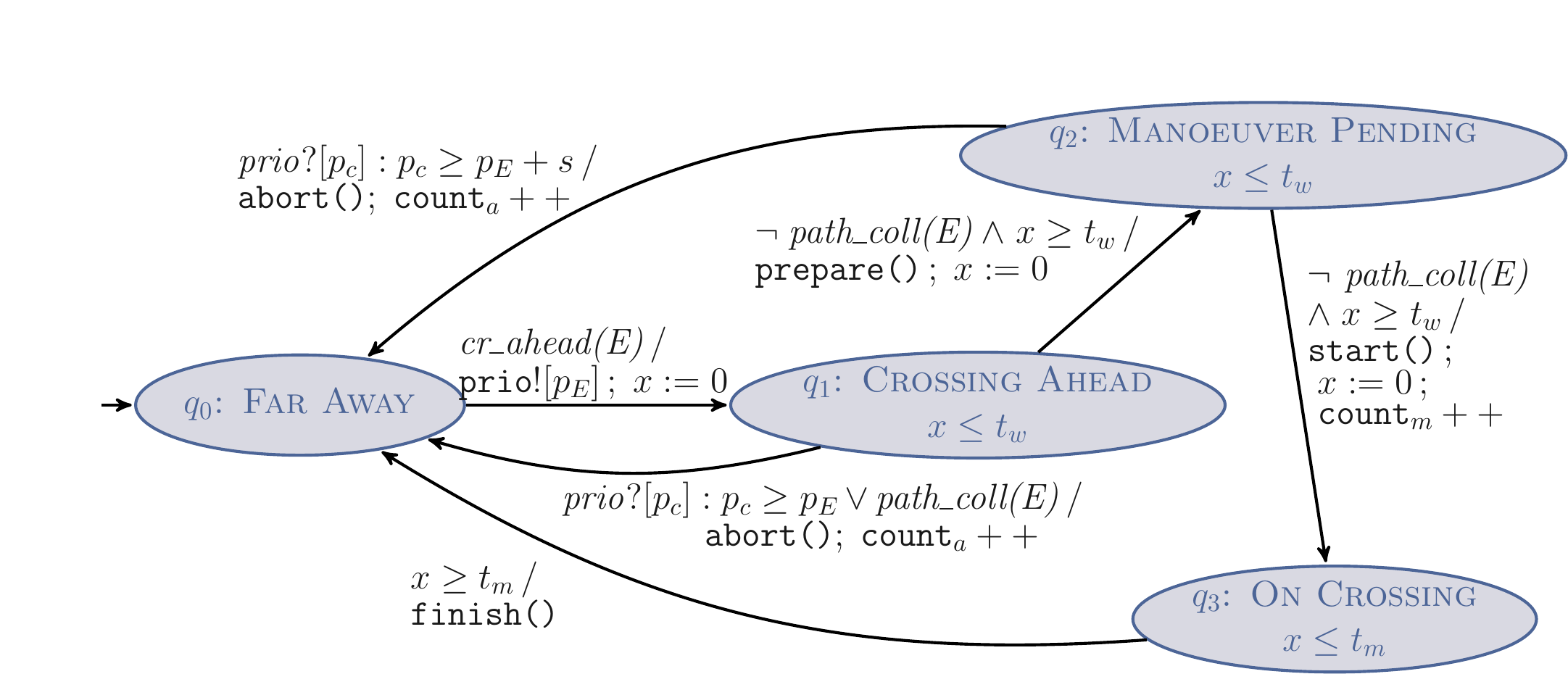}
    \caption{Simplified crossing controller protocol, inspired by \cite{BS19}.}
    \label{fig:case-study:protocol}
\end{figure}

We depict our simplified crossing protocol in Fig.~\ref{fig:case-study:protocol} and explain it in the following. The goal of this protocol is that an autonomous car, like the \ego{} car $E$ from Fig.~\ref{fig:pre:intersection}, can safely turn at an intersection. The protocol also implements a fairness property, where only the car with the highest priority may enter an intersection. Each car $C$ starts with a priority $p_c = 0$ and the longer a car waits, the more this priority increases. 

The protocol from Fig.~\ref{fig:case-study:protocol} summarises four phases, where each phase relates to one location $q_0$ to $q_3$ in the depicted timed automaton:
\begin{enumerate}
    \item \textsc{$q_0$ Far Away}: No crossing is in range.
    \item \textsc{$q_1$ Crossing Ahead:} On approaching a crossing ($\mathit{cr\_ahead()}$), the \ego{} car $E$ sends its own priority $p_E$ for entering via broadcast ($\mathit{prio!}[p_{E}]$) and compares it with the priorities of other traffic participants, and potential path collisions $\mathit{(path\_coll(E)}$) are checked.
    \item \textsc{$q_2$ Manoeuvre Pending: } The \ego{} car $E$ determines that its priority is the highest and that the desired path is free ($\neg \mathit{path\_coll(E)}$) before entering this phase. If $E$'s priority $p_E$ were smaller than the priority $p_c$ of an arbitrary car $C$, the guarded input action $\mathit{prio?}[p_{c}]: p_c \geq p_{E}$ on the transition from location $q_1$ to $q_0$ would be valid and $q_1$ would be left for $q_0$.
    \item \textsc{$q_3$ On Crossing:} The crossing is only entered if $E$ has the highest priority $p_E$ and no potential path collisions have been detected ($\neg \mathit{path\_coll(E)}$).
\end{enumerate}
Note that in phase 3 of the protocol, $q_2$ \textsc{Manoeuvre Pending}, the \ego{} car $E$ is about to enter the intersection, as its priority is higher than any other cars' priorities and as no collisions with its planned path through the intersection have been identified. The only possibility for \ego{} to abort its manoeuvre in this phase of the protocol is if a car with a significantly larger priority $p_c$ with $p_c \geq p_E + s$ arrives, e.g. an emergency vehicle. Such an emergency vehicle would not start with a priority of $0$ on arriving at the intersection, but with a much larger initial priority value.

In the timed automaton crossing protocol of Fig.~\ref{fig:case-study:protocol}, we observe three different types of \emph{actions}: 
\begin{itemize}
    \item Communication actions: With $\mathit{prio![p_{E}]}$, $E$'s priority $p_{E}$ is sent via broadcast to all other cars.
    \item Controller actions: Preparing ($\mathtt{prepare()}$), aborting ($\mathtt{abort()}$), starting ($\mathtt{start()}$) and finishing ($\mathtt{finish()}$) a crossing manoeuvre.
    \item Operations on data variables and clock resets: E.g., $x:=0$ to reset the value of a clock variable $x$ or \texttt{count}$_m$ (resp. \texttt{count}$_a$) to increase a counter for started (resp. aborted) manoeuvres.
\end{itemize}
The controller actions are a construct that has been defined for the special type of extended timed automata, Automotive-Controlling Timed Automata, in \cite{Sch18-TCS}, but can be understood as abstract functions for our purposes in this paper.
To model that actions like $\mathtt{start()}$ do not happen immediately, we add a time invariant, combined with a respective guard on the outgoing transitions, to most locations of the protocol. For instance consider the invariant $x \leq t_w$ in location $q_1$, which specifies that the location must be left after at most $t_w$ time units. Combined with the guard $x\geq t_w$ on the outgoing transition to $q_2$, this transition can only be taken after exactly $t_w$ time units. Location $q_1$ 
can only be left for location $q_0$ earlier than $t_w$ time units, if either a path collision was detected with the guard  $\mathit{path\_coll()}$ or if a higher priority $p_c$ has been received and identified via the \emph{guarded input action} $\mathit{prio?[p_{c}]}: p_c \geq p_{E}$. The communication semantics of timed automata specifies that, if a communication via the \emph{output action} $\mathit{prio![p_{C}]}$ has been received from another car $C$'s timed automaton controller, and if the communication guard $p_c \geq p_{E}$ holds for a variable valuation $\nu(c)=C$ of the variable $c$, the ego controller \textit{must} synchronise with this communication and thus change back to location $q_0$.
For more details on this type of guarded broadcast communication, we refer to \cite{Sch18-TCS}.

\section{Phase 1: Extraction and Preparation}\label{sec:model-extraction}
The first phase of our framework in Fig.~\ref{fig:ex-models} contains two steps, which we describe separately: First (Sect.~\ref{sec:model-extract}), the \exmodel{} is extracted from the \sysmodel{} and then (Sect.~\ref{sec:model-preparation}) the extracted model is prepared for the next phase of the framework. In each section, we first describe the respective step in general, and after that we apply the step to our running example (cf.~Sect.~\ref{sec:case-study}).

\subsection{Extraction}\label{sec:model-extract}
In this step, we extract a first version \modelone{} of the \exmodel{} from the \sysmodel{} \sys{}. For this, we require the following assumptions to hold for our system model:
\begin{enumerate}
    \item Completeness: Each possible system behaviour is modelled within the \sysmodel{} \sys.
    \item Correctitude: The \sysmodel{} is required to accurately and correctly represent the system's behaviour.
\end{enumerate}
If the first assumption is not satisfied, we can still extract an explanation model but this then only allows for partially explaining the system behaviour. Equally, if the second assumption is not satisfied, we can also still extract an explanation model, but it most likely would not be correct. With such a faulty \exmodel, we could then apply debugging methods to actually eliminate incorrect behaviour from the system model itself through faulty explanations. However, for our example, we assume that both assumptions hold to extract a correct explanation model. 

For now, we focus on timed automata \sysmodel s. For other types of \sysmodel s, the extraction process may differ, as we discuss in Sect.~\ref{sec:evaluation}.

For the extraction, we readopt the idea of \cite{Sch21-quest} to connect system actions with their reasons. However, we broaden the idea of actions to \emph{observables}. Such an observable may be any behaviour that can be observed from the outside of the system. We assume that a communication action or setting an internal variable to a new value is also an observable, as, e.g., an engineer or another system might be interested to keep track of the valuation of certain system variables.
Possible \emph{reasons} basically are the preconditions for an observable. Note that the term ``observable'' does not only comprise those elements that are ``visible'' for a human. Instead, an observable in our sense is some observable system behaviour, which could, e.g., be identified via observer automata, as they are generally used for timed automata.

To retrieve the reasons for an observable, a backwards search through our system model can find all possible \expath s that may lead to an observable. On assembling these \expath s to one model, we get a causal tree structure that is structurally comparable to the causal diagram that has been described in \cite{Sch21-quest}. The merits of such a tree structure are that on observing a certain phenomenon, the potential reasons and explanations can be directly extracted from this causal tree structure. However, we discuss potential limits of using a causal structure in Sect.~\ref{sec:evaluation}.

\textit{Running Example: }
In our running example, we assume that our explanation purpose is to explain unexpected behaviour of an AV at an intersection, and that the explainee is a passenger of the AV.\label{page:passenger}

To connect observables with their reasons, we exploit the extended timed automata semantics of our crossing protocol and connect \emph{actions} in our timed automaton model with all their possible \emph{reasons} (cf.~Sect.~\ref{sec:case-study}).
This idea of connecting actions with their reasons (resp. causes) stems from the ``models of causality'' approach \cite{Getal18}. 
%
A specific example for an action that we use in our protocol is the communication action $\mathit{prio!}[p_{E}]$, with which the \ego{} car communicates its priority $\mathit{p_{E}}$ for turning at the intersection (cf.~Fig.~\ref{fig:case-study:protocol}). A guard (i.e. reason) preceding this action is the function $\mathit{cr\_ahead()}$.
An example of an explanation path would connect the action $\mathit{prio!}[p_{ego}]$ with the reason $\mathit{cr\_ahead()}$.

\begin{figure}[ht]
\vspace{-0.0cm}
    \centering
    \includegraphics[width=\linewidth]{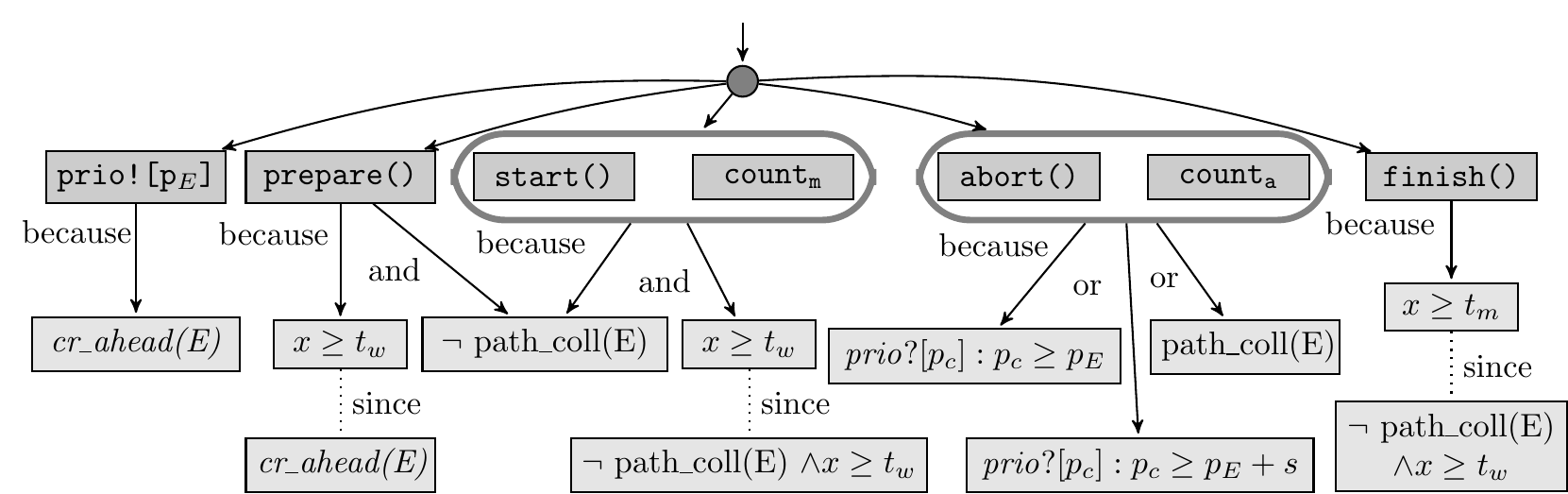}
    \vspace{-0.0cm}
    \caption{Explanation model \modelone{} connects all possible actions of the \sysmodel{} \sys{} with their possible reasons.} 
    \label{fig:model1}
    \vspace{-0.0cm}
\end{figure}

Note that for this example we do not include clock reset actions $x:=0$ for a clock variable $x$ into our notion of observables. This is because we assume that resetting of clock variables is an internal concept of timed automata.
The explanation model that we extract by connecting actions with their reasons is depicted in Fig,~\ref{fig:model1}. Note that this first version \modelone{} contains reasons for all observables. Also note that the superstates which we use in Fig.~\ref{fig:model1} are only used to simplify the figure: Their meaning is that, e.g., both the observables \texttt{count}$_a$ and \texttt{start()} have the same reasons. 

\subsection{Preparation}\label{sec:model-preparation}
In our second step of phase 1, we hide context details that are irrelevant for our explanation purpose. 
Thus, we reduce the content of our explanation model \modelone{} to behaviour that actually fits the explanation purpose within \modeltwo.

In our meaning, \emph{relevant information} is such information that helps the explainee to better understand system behaviour and functionality, w.r.t. an explanation purpose. To formally capture this notion of \emph{understanding explanations}, related approaches \cite{FFD22, BHRS22} compare the actual world model $W$ with a locally believed model $M_i$ of an explainee $i$. A relevant explanation is one that helps in transforming a believed model $M_i$ closer to the actual world model $W$.
As an example for irrelevant information, we refer back to our factory robot from Sect.~\ref{sec:creation-process}, p.~\pageref{sec:creation-process}.
Note that we do not yet hide information that are irrelevant for specific explainee types in this step. For this we refer to Sect.~\ref{sec:model-tailor}.

To reduce the content of \exmodel{} \modelone, we hide entire branches: Those that contain actions that are not connected to our explanation purpose. However, we do not add or remove, nor alter in any way, information within any of the other branches. With this, the suggested procedure is a variation of program slicing, which was introduced in \cite{Weiser81}, where the explanation purpose is used as slicing criterion. For future work, we want to investigate 
more sophisticated slicing methods, were we do not necessarily hide entire branches in the \exmodel.

\begin{figure}[ht]
    \centering
    \includegraphics[width=\linewidth]{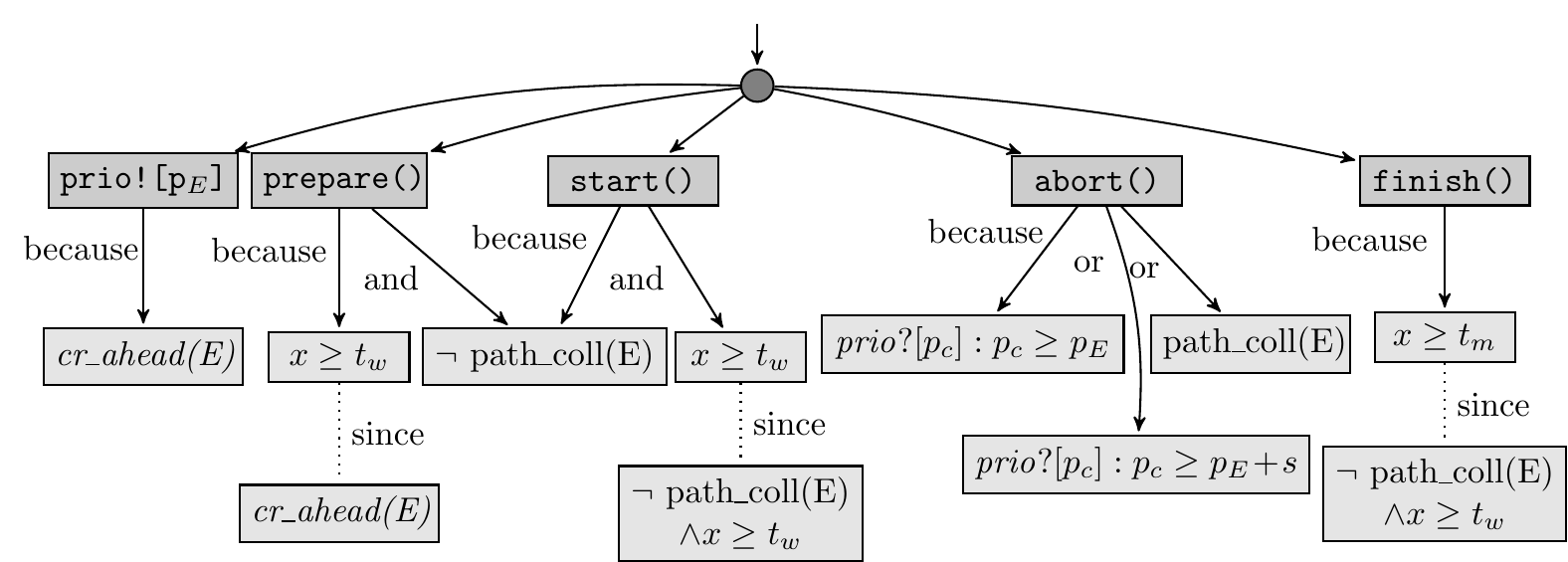}
    \caption{In \exmodel{} \modeltwo, branches that are related to actions out of the scope of the explanation purpose are removed from \modelone{} using a variation of slicing.}
    \label{fig:model2}
\end{figure}

\textit{Running Example: }
Consider again our running example, where we stated in Sect.~\ref{sec:model-extract} that our explainee is a passenger (cf. end-user) of an AV. We assume that internal system variables are not of interest for explaining visible behaviour of the AV to a passenger. Thus, we hide those actions that assign new values to data variables. I.e., we hide all branches that relate to the actions $\mathtt{count_a}$ and $\mathtt{count_m}$. This means that the superstates containing multiple actions in \modelone{} are obsolete in \modeltwo. The resulting \exmodel{} \modeltwo{} is depicted in Fig.\ref{fig:model2}.

\section{Phase 2: Tailoring and Refinement}\label{sec:model-refinement}
This second phase of our framework from Fig.~\ref{fig:ex-models} contains again two steps, which we again describe separately: First (Sect.~\ref{sec:model-tailor}, the \exmodel{} is tailored towards specific explainees and then (Sect.~\ref{sec:model-add-information}) context information is added. 
After this phase, the \exmodel{} may be deployed. Note that we again explain our concepts by using our running example. 

\subsection{Tailoring Towards Explainee}\label{sec:model-tailor}
Recall that before, in Sect.~\ref{sec:model-preparation}, we suggested to hide information that are irrelevant \emph{w.r.t. an explanation purpose}. In this step, we use another notion of relevance: Relevance w.r.t. a specific explainee type.

Thus, the goal of this step is to tailor the \exmodel{} \modeltwo{} towards a specific type of explainee, e.g. to an end-user, an engineer or another system. 
For this, explainee-specific details of the model may be eliminated. The intuition is that an end-user may not be interested in internal data variables, while an engineer might very well need explanations for values of such variables.
We again suggest to use slicing methods to reduce the \exmodel{} \modeltwo{} to a user-type-specific \exmodel{} \modelthree. We again hide entire branches that contain reasons for actions out of the scope for a specific explainee.

\textit{Running Example: }
We continue our case-study, where we consider that our explainee is an end-user and that the explanation purpose is to explain the behaviour of the crossing controller from Fig.~\ref{fig:case-study:protocol}.
We assume that an end-user will only be interested in explanations for visible behaviour of the AV, not perceiving the invisible behaviour. Thus, we omit the explanation branch for the action $\mathtt{prio!}[p_{E}]$, where the AV communicates its priority to other traffic participants. Further on, we assume that both the actions $\mathtt{prepare()}$ and $\mathtt{finish()}$ are not of interest for an end-user: the manoeuvre preparation comprises an internal system state, which is not visible to an end-user and while finishing the manoeuvre is visible (``leaving the intersection''), from the perspective of an end-user, this might not be perceived as an action that needs to be explained.

\begin{figure}[ht]
\vspace{-0.0cm}
    \begin{center}
    \includegraphics[width=0.8\linewidth]{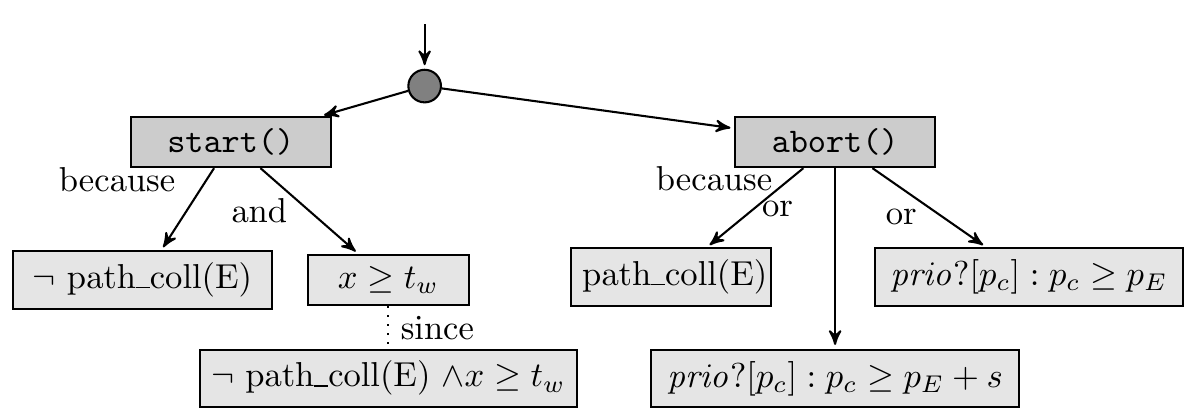}
    \caption{For tailoring \exmodel{} \modeltwo{} towards an end-user as explainee, invisible, internal system actions are omitted, resulting in a smaller \exmodel{} \modelthree.}
    \label{fig:model3}
    \end{center}
    \vspace{-0.0cm}
\end{figure}

Thus, the resulting \exmodel{} \modelthree{} that we depict in Fig.~\ref{fig:model3} only comprises the explanation branches for the actions $\mathtt{start()}$ and $\mathtt{abort()}$. Especially the latter one is of interest, as it is a version of a ``why not'' explanation (``Why do we not start the manoeuvre?''). Such explanations have been found to be of interest in \cite{LDA09}.

\subsection{Refinement: Add environmental information}\label{sec:model-add-information}
In this step, the \exmodel{} \modelthree{} is extended by explainee-specific environmental information. The intuition is that some of the information that is needed for a good explanation might not be contained within the \sysmodel{} \sys. Such information could be rules that the system abides by that are only encoded by abstract data variables within \sys. These information can be obtained by different sources: for example, from an engineer or from system requirements. If the addressee is an end-user or non-expert, we also suggest that abstract system data is annotated with natural language text snippets in this step. Note that these text snippets are not yet fragments of a formulated explanation that is provided to an addressee but should rather be understood as modules from which an explanation may be retrieved later on. We further clarify this progress from \modelthreex{} to \modelfourx{} with our running example in the following.

\textit{Running Example: }
In the case of our end-user, we suggest to annotate \modelthree{} with text snippets that can later be used to formulate explanations. For instance, the abstract clock variable $x$ and it exceeding some constant $t_w$ has no meaning to an end-user. However, if the right-hand branch of the action \texttt{start()} is summarised by ``manoeuvre time exceeded'', this text snipped might be integrated into an explanation later on. Further on, consider the reasons for \texttt{abort()}: A special case is the reason where the priority $p_c$ of another AV exceeds the priority $p_E$ of our AV $E$ under consideration \emph{significantly} (by a constant value ``s''). While this value ``s'' is abstract and without any meaning for an end-user, its meaning reflects the traffic rule where an emergency vehicle has the right of way at an intersection. 
This again is a user-understandable, and moreover, user-acceptable information.
\begin{figure}[ht]
\vspace{-0.0cm}
    \centering
    \includegraphics[width=\linewidth]{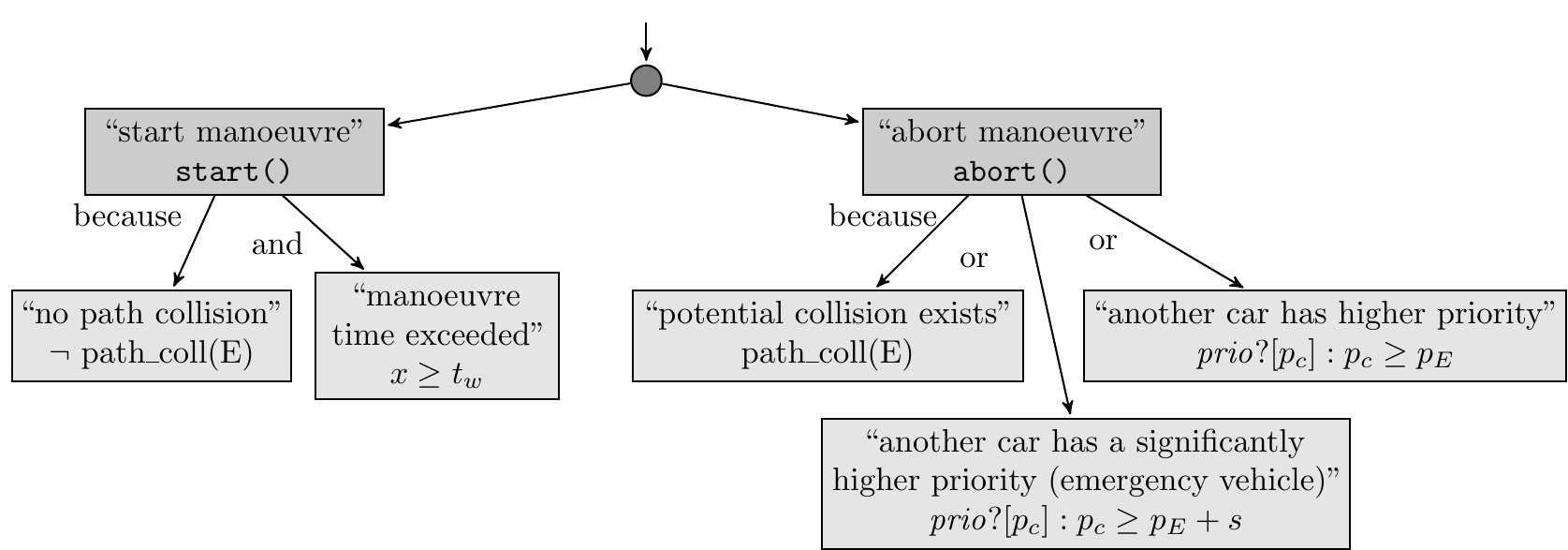}
    \vspace{-0.0cm}
    \caption{Explanation model \modelfour{} is enriched by natural language annotations to simplify the generation of explanations from this model later on. Further on, information about traffic rules, for instance that emergency vehicles have a right of way (``having the highest priority'') is reflected within \modelfour.}
    \label{fig:model3x}
\end{figure}

\section{Phase 3: Run-time Adaptation}\label{sec:updates}

We consider the \exmodel{} \modelfourx{} from the previous phase to be ready for deployment for the explainee type $X$. However, we postulate that an \exmodel{} should be individualised towards a specific explainee $x \in X$. Thus, on deploying the \exmodel, we consider an initial \exmodel{} \modelfivexone{} with $\modelfivexone\sim\modelfourx$ for an individual explainee $x_1 \in X$.
For different explainees $x_1, x_2$, a further individualisation or personalisation of the initial models \modelfivexone, \modelfivextwo{} is necessary as different explainees have different explanation preferences, prior system knowledge and attention levels that may influence their cognitive workload. Such different attention levels have, e.g., been identified for air-plane pilots within the \emph{Salience, Effort, Expectancy and Value (SEEV)} model in \cite{Wickens01}. Also, users gain experience with the system during its lifetime and might want to hide specific (types of) explanations on demand.
Equally, a run-time adaptation of \modelfivexone{} will be necessary, whenever the \sysmodel{} itself changes, e.g. due to self-adaptation, learning or software updates.  
We identify possibilities to detect the need for model updates in the following and leave the techniques to update the \exmodel{} for future work.

\subsubsection*{System Update-triggered Adaptations. }
The need for model updates of \modelfivexone{} due to an updated \sysmodel{} can be automatically identified if a corresponding update process for the \sysmodel{} exists. With this updated model, we can start the extraction and refinement process again. To do this automatically at run-time, all steps need to be fully automated. However, to reduce the computation overhead, we plan to investigate an incremental approach that adds or removes parts of the \exmodel s in future work. 
 This requires a meta model for the \exmodel{} that defines operations with which the \exmodel{} can be adjusted. This could, for example, be realised by graph transformation rules.

During run-time, the explanation layer could also detect the need for updating the \sysmodel{} when an explanation cannot be extracted from the existing \exmodel{} \modelfivex, e.g. due to a novel situation that the system encounters.

\subsubsection*{User-centered Adaptations. }
We identify several possibilities to detect the need for 
those updates that are necessary for the further individualisation towards the explainee $x_1$:
\begin{itemize}
    \item The explainee $x_1$ requests a more detailed explanation that cannot be extracted from the existing \exmodel{} \modelfivex
    \item The explainee $x_1$ requests to hide a specific explanation 
    \item Through observation of the user's behaviour, the system detects that they have gained experience with the system and do not need detailed explanations for known situations.
\end{itemize}
The first case is connected to findings that ``explaining'' is not a static process, where an explanation is given only once. Instead, related work \cite{KA97} suggests that not only one isolated explanation, but instead an \emph{explanation process} should be considered. In such a process, the system may start with a brief explanation and refine this explanation later, if the user is not initially satisfied with it. For this, a differently detailed \exmodel{} might be needed, for which the different phases of our approach are ideal. We will consider such dynamic explanation processes in future work.
To enable adjustments for specific explainees at run-time, we assume that the explainee can give feedback for each explanation, e.g., by rating whether the explanation was helpful and whether the level of detail was adequate.




\section{Discussion and Future Work}\label{sec:evaluation}
In this paper, we introduce an extraction and refinement process for \exmodel s that allows making existing systems self-explainable by using an additional explanation layer like it is done in our MAB-EX framework. We showcase the usability of our approach with a timed automata running example. Nonetheless, our process is only the first stepping stone towards an automatic, generalised process of deriving explanations directly from \sysmodel s. Here, we discuss and assess the main strengths of our approach, but also name its limits and potential topics for future work.

\subsubsection*{Running Example and Generality.}
We have chosen a timed automaton specification (\sys) for our running example, as timed automata allow to formalise a good variety of system types. Also, for our specific example, a formal semantics and proofs for system properties like safety, already exist in \cite{Sch18-TCS, BS19}. With our running example, we enrich \cite{BS19} by an explainability module. In future work, we plan to examine whether, for timed automata, some state history should also be integrated into the \exmodel{} \modelone.

Note that our extraction and refinement process from Fig.\ref{fig:ex-models} itself is not limited to timed automata. Indeed, our general framework is capable of retrieving \exmodel s from arbitrary types of \sysmodel s.
To use our approach for other types of systems, we currently identify an important additional requirement for the \sysmodel{} \sys{} (apart from those that were introduced in the beginning of Sect.~\ref{sec:model-extract}): It must be possible to identify observables (``actions'') and reasons within the system. We intend to examine whether this additional requirement indeed is enough with further case-studies. Also, a generalised definition for the \exmodel{} extraction process for causality-based \sysmodel s should be provided.

\subsubsection*{Automation.}
Most steps of our approach do not require human intervention and could be done automatically. In future work, we thus aim to formally define the requirements for all steps of our process and to implement them to automatically extract an \exmodel{} from a \sysmodel. To this end, we will also investigate elaborate slicing techniques to tailor the extracted \exmodel{} to different stakeholders.
Only for the tailoring of \modelthreex{} to \modelfourx, we suggest to involve an expert, e.g. to connect internal system data to rules. For future work, we intend to examine whether such information could also be automatically retrieved from other sources. While currently, we only use the \sysmodel{} \sys{} for constructing our \exmodel, one could also exploit other available sources from the entire system development process: For instance, a requirements document or system code and its code documentation might be of help to annotate abstract observables and reasons with text fragments (cf. as in Sect.~\ref{sec:model-add-information}).

\subsubsection*{Integration into MAB-EX.}
Following the MAB-EX framework, the \exmodel{} that we present here is not intended to generate actual explanations that would be presented to an end-user. Instead, we retrieve internal explanation paths from our \exmodel{} in the \emph{Build} phase. The next step would be to translate these internal explanation paths into actual explanations in the last MAB-EX phase. For this, e.g., \cite{R11} discusses translation problems between language and logical representations and a variety of approaches exists that discuss how to best present explanations.

\subsubsection*{Formal Models for Explainability.}
Our \exmodel{} is a formal model for explanations, as it is retrieved from a formal \sysmodel. 
%
With that, we have internal, formalised explanation paths that allow for a translation for different types of explainees, e.g. experts or other systems. Also, our explanation paths are concise, formal representations of explanations. With this, information loss due to natural language translation is minimised.
Further on, we envision that we can actually formally analyse and prove certain properties of explanations in future work. For this, our notion of explanation paths would need to be integrated into a formal definition of explanations.



\section{Conclusion}\label{sec:conclusion}
We present a high-level process for extracting and refining \exmodel s from formal \sysmodel s, thus allowing to make an existing system self-explainable. From our \exmodel, explanation paths may be automatically retrieved at run-time and then be translated into explainee-understandable explanations. A strength of our approach is that we take different types of explainees into account: e.g. an end-user, an engineer, or another system. Further on, our formal approach to explanations ensures that future formal analyses are possible. Also, not only the final \exmodel{} \modelfivexone{} is of use for the process of explaining system behaviour. The different intermediate \exmodel s from each of our phases themselves show a different degree of detail which could be used to integrate these models into other approaches. We showcase our approach using a case-study from the automotive domain, where we extract and refine an \exmodel{} from an extended timed automaton controller for turn manoeuvres at intersections.
While our reference process is a first step towards a system capability of automatic self-explanation of system behaviour, we identify and sketch further tasks in our future work section.

\bibliographystyle{eptcs}
\bibliography{bib}

\begin{thebibliography}{10}
\providecommand{\bibitemdeclare}[2]{}
\providecommand{\surnamestart}{}
\providecommand{\surnameend}{}
\providecommand{\urlprefix}{Available at }
\providecommand{\url}[1]{\texttt{#1}}
\providecommand{\href}[2]{\texttt{#2}}
\providecommand{\urlalt}[2]{\href{#1}{#2}}
\providecommand{\doi}[1]{doi:\urlalt{http://dx.doi.org/#1}{#1}}
\providecommand{\bibinfo}[2]{#2}

\bibitemdeclare{techreport}{ibm2005architectural}
\bibitem{ibm2005architectural}
 (\bibinfo{year}{2005}): \emph{\bibinfo{title}{An Architectural Blueprint for
  Autonomic Computing}}.
\newblock \bibinfo{type}{White Paper}, \bibinfo{institution}{IBM}.

\bibitemdeclare{inproceedings}{agrawal2021explaining}
\bibitem{agrawal2021explaining}
\bibinfo{author}{Ankit \surnamestart Agrawal\surnameend} \&
  \bibinfo{author}{Jane \surnamestart Cleland-Huang\surnameend}
  (\bibinfo{year}{2021}): \emph{\bibinfo{title}{Explaining Autonomous Decisions
  in Swarms of Human-on-the-Loop Small Unmanned Aerial Systems}}.
\newblock In: {\sl \bibinfo{booktitle}{Proceedings of the AAAI Conference on
  Human Computation and Crowdsourcing}}, \bibinfo{volume}{9}, pp.
  \bibinfo{pages}{15--26}.
\newblock
  \urlprefix\url{https://ojs.aaai.org/index.php/HCOMP/article/view/18936}.

\bibitemdeclare{inproceedings}{BHRS22}
\bibitem{BHRS22}
\bibinfo{author}{Astrid~Rakow \surnamestart Akhila~Bairy\surnameend,
  Willem~Hagemann} \& \bibinfo{author}{Maike \surnamestart
  Schwammberger\surnameend} (\bibinfo{year}{2022}):
  \emph{\bibinfo{title}{Towards formal concepts for explanation timing and
  justifications}}.
\newblock In: {\sl \bibinfo{booktitle}{2022 IEEE 30th International
  Requirements Engineering Conference Workshops (REW)}}.

\bibitemdeclare{article}{AD94}
\bibitem{AD94}
\bibinfo{author}{Rajeev \surnamestart Alur\surnameend} \&
  \bibinfo{author}{David~L. \surnamestart Dill\surnameend}
  (\bibinfo{year}{1994}): \emph{\bibinfo{title}{A Theory of Timed Automata}}.
\newblock {\sl \bibinfo{journal}{Theoretical Computer Science}}
  \bibinfo{volume}{126}(\bibinfo{number}{2}), pp. \bibinfo{pages}{183--235},
  \doi{10.1016/0304-3975(94)90010-8}.

\bibitemdeclare{inproceedings}{anjomshoae2019explainable}
\bibitem{anjomshoae2019explainable}
\bibinfo{author}{Sule \surnamestart Anjomshoae\surnameend},
  \bibinfo{author}{Amro \surnamestart Najjar\surnameend},
  \bibinfo{author}{Davide \surnamestart Calvaresi\surnameend} \&
  \bibinfo{author}{Kary \surnamestart Fr{\"a}mling\surnameend}
  (\bibinfo{year}{2019}): \emph{\bibinfo{title}{Explainable agents and robots:
  Results from a systematic literature review}}.
\newblock In: {\sl \bibinfo{booktitle}{18th International Conference on
  Autonomous Agents and Multiagent Systems (AAMAS 2019), Montreal, Canada, May
  13--17, 2019}}, \bibinfo{organization}{International Foundation for
  Autonomous Agents and Multiagent Systems}, pp. \bibinfo{pages}{1078--1088},
  \doi{10.5555/3306127.3331806}.

\bibitemdeclare{inproceedings}{BS19}
\bibitem{BS19}
\bibinfo{author}{Christopher \surnamestart Bischopink\surnameend} \&
  \bibinfo{author}{Maike \surnamestart Schwammberger\surnameend}
  (\bibinfo{year}{2019}): \emph{\bibinfo{title}{Verification of Fair
  Controllers for Urban Traffic Manoeuvres at Intersections}}.
\newblock In \bibinfo{editor}{Emil \surnamestart Sekerinski\surnameend},
  \bibinfo{editor}{Nelma \surnamestart Moreira\surnameend},
  \bibinfo{editor}{Jos{\'{e}}~N. \surnamestart Oliveira\surnameend},
  \bibinfo{editor}{Daniel \surnamestart Ratiu\surnameend},
  \bibinfo{editor}{Riccardo \surnamestart Guidotti\surnameend},
  \bibinfo{editor}{Marie \surnamestart Farrell\surnameend},
  \bibinfo{editor}{Matt \surnamestart Luckcuck\surnameend},
  \bibinfo{editor}{Diego \surnamestart Marmsoler\surnameend},
  \bibinfo{editor}{Jos{\'{e}} \surnamestart Campos\surnameend},
  \bibinfo{editor}{Troy \surnamestart Astarte\surnameend},
  \bibinfo{editor}{Laure \surnamestart Gonnord\surnameend},
  \bibinfo{editor}{Antonio \surnamestart Cerone\surnameend},
  \bibinfo{editor}{Luis \surnamestart Couto\surnameend},
  \bibinfo{editor}{Brijesh \surnamestart Dongol\surnameend},
  \bibinfo{editor}{Martin \surnamestart Kutrib\surnameend},
  \bibinfo{editor}{Pedro \surnamestart Monteiro\surnameend} \&
  \bibinfo{editor}{David \surnamestart Delmas\surnameend}, editors: {\sl
  \bibinfo{booktitle}{Formal Methods. {FM} 2019 International Workshops -
  Porto, Portugal, October 7-11, 2019, Revised Selected Papers, Part {I}}},
  {\sl \bibinfo{series}{Lecture Notes in Computer Science}}
  \bibinfo{volume}{12232}, \bibinfo{publisher}{Springer}, pp.
  \bibinfo{pages}{249--264}, \doi{10.1007/978-3-030-54994-7\_18}.

\bibitemdeclare{inproceedings}{Betal19}
\bibitem{Betal19}
\bibinfo{author}{Mathias \surnamestart Blumreiter\surnameend},
  \bibinfo{author}{Joel \surnamestart Greenyer\surnameend},
  \bibinfo{author}{Francisco Javier~Chiyah \surnamestart Garcia\surnameend},
  \bibinfo{author}{Verena \surnamestart Kl{\"{o}}s\surnameend},
  \bibinfo{author}{Maike \surnamestart Schwammberger\surnameend},
  \bibinfo{author}{Christoph \surnamestart Sommer\surnameend},
  \bibinfo{author}{Andreas \surnamestart Vogelsang\surnameend} \&
  \bibinfo{author}{Andreas \surnamestart Wortmann\surnameend}
  (\bibinfo{year}{2019}): \emph{\bibinfo{title}{Towards Self-Explainable
  Cyber-Physical Systems}}.
\newblock In: {\sl \bibinfo{booktitle}{22nd {ACM/IEEE} International Conference
  on Model Driven Engineering Languages and Systems Companion}}, pp.
  \bibinfo{pages}{543--548}, \doi{10.1109/MODELS-C.2019.00084}.

\bibitemdeclare{inproceedings}{Getal18}
\bibitem{Getal18}
\bibinfo{author}{Francisco~Javier \surnamestart Chiyah~Garcia\surnameend},
  \bibinfo{author}{David~A. \surnamestart Robb\surnameend},
  \bibinfo{author}{Xingkun \surnamestart Liu\surnameend},
  \bibinfo{author}{Atanas \surnamestart Laskov\surnameend},
  \bibinfo{author}{Pedro \surnamestart Patron\surnameend} \&
  \bibinfo{author}{Helen \surnamestart Hastie\surnameend}
  (\bibinfo{year}{2018}): \emph{\bibinfo{title}{Explainable Autonomy: A Study
  of Explanation Styles for Building Clear Mental Models}}.
\newblock In: {\sl \bibinfo{booktitle}{Proceedings of the 11th International
  Conference on Natural Language Generation}}, \bibinfo{publisher}{Association
  for Computational Linguistics}, \bibinfo{address}{Tilburg University, The
  Netherlands}, pp. \bibinfo{pages}{99--108}, \doi{10.18653/v1/W18-6511}.
\newblock \urlprefix\url{https://www.aclweb.org/anthology/W18-6511}.

\bibitemdeclare{book}{reqeng17}
\bibitem{reqeng17}
\bibinfo{author}{Jeremy \surnamestart Dick\surnameend},
  \bibinfo{author}{M.~Elizabeth~C. \surnamestart Hull\surnameend} \&
  \bibinfo{author}{Ken \surnamestart Jackson\surnameend}
  (\bibinfo{year}{2017}): \emph{\bibinfo{title}{Requirements Engineering, 4th
  Edition}}.
\newblock \bibinfo{publisher}{Springer}, \doi{10.1007/978-3-319-61073-3}.

\bibitemdeclare{inproceedings}{FLSM22}
\bibitem{FLSM22}
\bibinfo{author}{Marie \surnamestart Farrell\surnameend}, \bibinfo{author}{Matt
  \surnamestart Luckcuck\surnameend}, \bibinfo{author}{Ois{\'{\i}}n
  \surnamestart Sheridan\surnameend} \& \bibinfo{author}{Rosemary \surnamestart
  Monahan\surnameend} (\bibinfo{year}{2022}): \emph{\bibinfo{title}{FRETting
  About Requirements: Formalised Requirements for an Aircraft Engine
  Controller}}.
\newblock In \bibinfo{editor}{Vincenzo \surnamestart Gervasi\surnameend} \&
  \bibinfo{editor}{Andreas \surnamestart Vogelsang\surnameend}, editors: {\sl
  \bibinfo{booktitle}{Requirements Engineering: Foundation for Software Quality
  - 28th International Working Conference, {REFSQ} 2022, Birmingham, UK, March
  21-24, 2022, Proceedings}}, {\sl \bibinfo{series}{Lecture Notes in Computer
  Science}} \bibinfo{volume}{13216}, \bibinfo{publisher}{Springer}, pp.
  \bibinfo{pages}{96--111}, \doi{10.1007/978-3-030-98464-9\_9}.

\bibitemdeclare{inproceedings}{goebel2018explainable}
\bibitem{goebel2018explainable}
\bibinfo{author}{Randy \surnamestart Goebel\surnameend}, \bibinfo{author}{Ajay
  \surnamestart Chander\surnameend}, \bibinfo{author}{Katharina \surnamestart
  Holzinger\surnameend}, \bibinfo{author}{Freddy \surnamestart
  Lecue\surnameend}, \bibinfo{author}{Zeynep \surnamestart Akata\surnameend},
  \bibinfo{author}{Simone \surnamestart Stumpf\surnameend},
  \bibinfo{author}{Peter \surnamestart Kieseberg\surnameend} \&
  \bibinfo{author}{Andreas \surnamestart Holzinger\surnameend}
  (\bibinfo{year}{2018}): \emph{\bibinfo{title}{Explainable AI: the new 42?}}
\newblock In: {\sl \bibinfo{booktitle}{International cross-domain conference
  for machine learning and knowledge extraction}},
  \bibinfo{organization}{Springer}, pp. \bibinfo{pages}{295--303},
  \doi{10.1007/978-3-319-99740-7\_21}.

\bibitemdeclare{inproceedings}{FFD22}
\bibitem{FFD22}
\bibinfo{author}{Martin~Fränzle \surnamestart Goerschwin~Fey\surnameend} \&
  \bibinfo{author}{Rolf \surnamestart Drechsler\surnameend}
  (\bibinfo{year}{2022}): \emph{\bibinfo{title}{Self-Explanation in Systems of
  Systems}}.
\newblock In: {\sl \bibinfo{booktitle}{2022 IEEE 30th International
  Requirements Engineering Conference Workshops (REW)}}.

\bibitemdeclare{article}{greenyer2019explainable}
\bibitem{greenyer2019explainable}
\bibinfo{author}{Joel \surnamestart Greenyer\surnameend},
  \bibinfo{author}{Malte \surnamestart Lochau\surnameend} \&
  \bibinfo{author}{Thomas \surnamestart Vogel\surnameend}
  (\bibinfo{year}{2019}): \emph{\bibinfo{title}{Explainable software for
  cyber-physical systems (es4cps): Report from the gi dagstuhl seminar 19023,
  january 06-11 2019, schloss dagstuhl}}.
\newblock {\sl \bibinfo{journal}{arXiv:1904.11851}},
  \doi{10.48550/arXiv.1904.11851}.

\bibitemdeclare{inproceedings}{Holzinger2022}
\bibitem{Holzinger2022}
\bibinfo{author}{Andreas \surnamestart Holzinger\surnameend},
  \bibinfo{author}{Anna \surnamestart Saranti\surnameend},
  \bibinfo{author}{Christoph \surnamestart Molnar\surnameend},
  \bibinfo{author}{Przemyslaw \surnamestart Biecek\surnameend} \&
  \bibinfo{author}{Wojciech \surnamestart Samek\surnameend}
  (\bibinfo{year}{2020}): \emph{\bibinfo{title}{Explainable {AI} Methods - {A}
  Brief Overview}}.
\newblock In \bibinfo{editor}{Andreas \surnamestart Holzinger\surnameend},
  \bibinfo{editor}{Randy \surnamestart Goebel\surnameend},
  \bibinfo{editor}{Ruth \surnamestart Fong\surnameend}, \bibinfo{editor}{Taesup
  \surnamestart Moon\surnameend}, \bibinfo{editor}{Klaus{-}Robert \surnamestart
  M{\"{u}}ller\surnameend} \& \bibinfo{editor}{Wojciech \surnamestart
  Samek\surnameend}, editors: {\sl \bibinfo{booktitle}{xxAI - Beyond
  Explainable {AI} - International Workshop, Held in Conjunction with {ICML}
  2020, July 18, 2020, Vienna, Austria, Revised and Extended Papers}}, {\sl
  \bibinfo{series}{Lecture Notes in Computer Science}} \bibinfo{volume}{13200},
  \bibinfo{publisher}{Springer}, pp. \bibinfo{pages}{13--38},
  \doi{10.1007/978-3-031-04083-2\_2}.

\bibitemdeclare{inproceedings}{CDF22}
\bibitem{CDF22}
\bibinfo{author}{Louise~Dennis \surnamestart Joe~Collenette\surnameend} \&
  \bibinfo{author}{Michael \surnamestart Fisher\surnameend}
  (\bibinfo{year}{2022}): \emph{\bibinfo{title}{Advising Autonomous Cars about
  the Rules of the Road}}.
\newblock In: {\sl \bibinfo{booktitle}{Proceedings of the Fourth Workshop on
  Formal Methods for Autonomous Systems, {FMAS@SEFM'22}, 26th-27th of September
  2022}}, \bibinfo{series}{{EPTCS}}.

\bibitemdeclare{article}{Ketal15}
\bibitem{Ketal15}
\bibinfo{author}{Jeamin \surnamestart Koo\surnameend}, \bibinfo{author}{Jungsuk
  \surnamestart Kwac\surnameend}, \bibinfo{author}{Wendy \surnamestart
  Ju\surnameend}, \bibinfo{author}{Martin \surnamestart Steinert\surnameend},
  \bibinfo{author}{Larry \surnamestart Leifer\surnameend} \&
  \bibinfo{author}{Clifford \surnamestart Nass\surnameend}
  (\bibinfo{year}{2015}): \emph{\bibinfo{title}{Why did my car just do that?
  Explaining semi-autonomous driving actions to improve driver understanding,
  trust, and performance}}.
\newblock {\sl \bibinfo{journal}{International Journal on Interactive Design
  and Manufacturing (IJIDeM)}} \bibinfo{volume}{9}(\bibinfo{number}{4}), pp.
  \bibinfo{pages}{269--275}, \doi{10.1007/s12008-014-0227-2}.

\bibitemdeclare{article}{KA97}
\bibitem{KA97}
\bibinfo{author}{Douglas~S. \surnamestart Krull\surnameend} \&
  \bibinfo{author}{Craig~A. \surnamestart Anderson\surnameend}
  (\bibinfo{year}{1997}): \emph{\bibinfo{title}{The Process of Explanation}}.
\newblock {\sl \bibinfo{journal}{Current Directions in Psychological Science}}
  \bibinfo{volume}{6}(\bibinfo{number}{1}), pp. \bibinfo{pages}{1--5},
  \doi{10.1111/1467-8721.ep11512447}.

\bibitemdeclare{inproceedings}{Ketal19}
\bibitem{Ketal19}
\bibinfo{author}{Maximilian~A. \surnamestart Köhl\surnameend},
  \bibinfo{author}{Kevin \surnamestart Baum\surnameend},
  \bibinfo{author}{Markus \surnamestart Langer\surnameend},
  \bibinfo{author}{Daniel \surnamestart Oster\surnameend},
  \bibinfo{author}{Timo \surnamestart Speith\surnameend} \&
  \bibinfo{author}{Dimitri \surnamestart Bohlender\surnameend}
  (\bibinfo{year}{2019}): \emph{\bibinfo{title}{Explainability as a
  Non-Functional Requirement}}.
\newblock In: {\sl \bibinfo{booktitle}{2019 IEEE 27th International
  Requirements Engineering Conference (RE)}}, pp. \bibinfo{pages}{363--368},
  \doi{10.1109/RE.2019.00046}.

\bibitemdeclare{inproceedings}{LDA09}
\bibitem{LDA09}
\bibinfo{author}{Brian~Y. \surnamestart Lim\surnameend},
  \bibinfo{author}{Anind~K. \surnamestart Dey\surnameend} \&
  \bibinfo{author}{Daniel \surnamestart Avrahami\surnameend}
  (\bibinfo{year}{2009}): \emph{\bibinfo{title}{Why and Why Not Explanations
  Improve the Intelligibility of Context-Aware Intelligent Systems}}.
\newblock In: {\sl \bibinfo{booktitle}{Proceedings of the SIGCHI Conference on
  Human Factors in Computing Systems}}, \bibinfo{series}{CHI '09},
  \bibinfo{publisher}{Association for Computing Machinery},
  \bibinfo{address}{New York, NY, USA}, p. \bibinfo{pages}{2119–2128},
  \doi{10.1145/1518701.1519023}.

\bibitemdeclare{inproceedings}{plambeck2022}
\bibitem{plambeck2022}
\bibinfo{author}{Swantje \surnamestart Plambeck\surnameend},
  \bibinfo{author}{G{\"o}rschwin \surnamestart Fey\surnameend},
  \bibinfo{author}{Jakob \surnamestart Schyga\surnameend},
  \bibinfo{author}{Johannes \surnamestart Hinckeldeyn\surnameend} \&
  \bibinfo{author}{Jochen \surnamestart Kreutzfeldt\surnameend}
  (\bibinfo{year}{2022}): \emph{\bibinfo{title}{Explaining Cyber-Physical
  Systems Using Decision Trees}}.
\newblock In: {\sl \bibinfo{booktitle}{2nd International Workshop on
  Computation-Aware Algorithmic Design for Cyber-Physical Systems (CAADCPS)}},
  \doi{10.1109/CAADCPS56132.2022.00006}.
\newblock
  \urlprefix\url{https://conferences.computer.org/cpsiot/pdfs/CAADCPS2022-5TICzWIXsIbzy5ctgEfHPL/820100a003/820100a003.pdf}.

\bibitemdeclare{inproceedings}{R11}
\bibitem{R11}
\bibinfo{author}{Aarne \surnamestart Ranta\surnameend} (\bibinfo{year}{2011}):
  \emph{\bibinfo{title}{Translating between Language and Logic: What Is Easy
  and What Is Difficult}}.
\newblock In \bibinfo{editor}{Nikolaj \surnamestart Bj{\o}rner\surnameend} \&
  \bibinfo{editor}{Viorica \surnamestart Sofronie{-}Stokkermans\surnameend},
  editors: {\sl \bibinfo{booktitle}{Automated Deduction - {CADE-23} - 23rd
  International Conference on Automated Deduction, Wroclaw, Poland, July 31 -
  August 5, 2011. Proceedings}}, {\sl \bibinfo{series}{Lecture Notes in
  Computer Science}} \bibinfo{volume}{6803}, \bibinfo{publisher}{Springer}, pp.
  \bibinfo{pages}{5--25}, \doi{10.1007/978-3-642-22438-6\_3}.

\bibitemdeclare{inproceedings}{Sadeghi21}
\bibitem{Sadeghi21}
\bibinfo{author}{Mersedeh \surnamestart Sadeghi\surnameend},
  \bibinfo{author}{Verena \surnamestart Kl{\"o}s\surnameend} \&
  \bibinfo{author}{Andreas \surnamestart Vogelsang\surnameend}
  (\bibinfo{year}{2021}): \emph{\bibinfo{title}{Cases for Explainable Software
  Systems: Characteristics and Examples}}.
\newblock In: {\sl \bibinfo{booktitle}{IEEE 29th International Requirements
  Engineering Conference Workshops (REW)}}, pp. \bibinfo{pages}{181--187},
  \doi{10.1109/REW53955.2021.00033}.

\bibitemdeclare{article}{Sch18-TCS}
\bibitem{Sch18-TCS}
\bibinfo{author}{Maike \surnamestart Schwammberger\surnameend}
  (\bibinfo{year}{2018}): \emph{\bibinfo{title}{An abstract model for proving
  safety of autonomous urban traffic}}.
\newblock {\sl \bibinfo{journal}{Theoretical Computer Science}}
  \bibinfo{volume}{744}, pp. \bibinfo{pages}{143--169},
  \doi{10.1016/j.tcs.2018.05.028}.

\bibitemdeclare{inproceedings}{Sch21-quest}
\bibitem{Sch21-quest}
\bibinfo{author}{Maike \surnamestart Schwammberger\surnameend}
  (\bibinfo{year}{2021}): \emph{\bibinfo{title}{A Quest of Self-Explainability:
  When Causal Diagrams meet Autonomous Urban Traffic Manoeuvres}}.
\newblock In: {\sl \bibinfo{booktitle}{2021 IEEE 29th International
  Requirements Engineering Conference Workshops (REW)}}, pp.
  \bibinfo{pages}{195--199}, \doi{10.1109/REW53955.2021.00035}.

\bibitemdeclare{inproceedings}{Weiser81}
\bibitem{Weiser81}
\bibinfo{author}{Mark \surnamestart Weiser\surnameend} (\bibinfo{year}{1981}):
  \emph{\bibinfo{title}{Program Slicing}}.
\newblock In: {\sl \bibinfo{booktitle}{Proceedings of the 5th International
  Conference on Software Engineering}}, \bibinfo{series}{ICSE '81},
  \bibinfo{publisher}{IEEE Press}, p. \bibinfo{pages}{439–449},
  \doi{10.1109/TSE.1984.5010248}.

\bibitemdeclare{article}{weyns2021research}
\bibitem{weyns2021research}
\bibinfo{author}{Danny \surnamestart Weyns\surnameend}, \bibinfo{author}{Jesper
  \surnamestart Andersson\surnameend}, \bibinfo{author}{Mauro \surnamestart
  Caporuscio\surnameend}, \bibinfo{author}{Francesco \surnamestart
  Flammini\surnameend}, \bibinfo{author}{Andreas \surnamestart
  Kerren\surnameend} \& \bibinfo{author}{Welf \surnamestart
  L\"{o}we\surnameend} (\bibinfo{year}{2021}): \emph{\bibinfo{title}{A Research
  Agenda for Smarter Cyber-Physical Systems}}.
\newblock {\sl \bibinfo{journal}{J. Integr. Des. Process Sci.}}
  \bibinfo{volume}{25}(\bibinfo{number}{2}), p. \bibinfo{pages}{27–47},
  \doi{10.3233/JID210010}.

\bibitemdeclare{inproceedings}{Wickens01}
\bibitem{Wickens01}
\bibinfo{author}{Christopher~D. \surnamestart Wickens\surnameend},
  \bibinfo{author}{John \surnamestart Helleberg\surnameend},
  \bibinfo{author}{Juliana \surnamestart Goh\surnameend},
  \bibinfo{author}{Xidong \surnamestart Xu\surnameend} \&
  \bibinfo{author}{William~J. \surnamestart Horrey\surnameend}
  (\bibinfo{year}{2001}): \emph{\bibinfo{title}{Pilot Task Management : Testing
  an Attentional Expected Value Model of Visual Scanning}}.
\newblock
  \urlprefix\url{http://apps.usd.edu/coglab/schieber/psyc792/workload/Wickens-etal-2001.pdf}.

\bibitemdeclare{inproceedings}{Ziesche21}
\bibitem{Ziesche21}
\bibinfo{author}{Florian \surnamestart Ziesche\surnameend},
  \bibinfo{author}{Verena \surnamestart Kl{\"o}s\surnameend} \&
  \bibinfo{author}{Sabine \surnamestart Glesner\surnameend}
  (\bibinfo{year}{2021}): \emph{\bibinfo{title}{Anomaly Detection and
  Classification to enable Self-Explainability of Autonomous Systems}}.
\newblock In: {\sl \bibinfo{booktitle}{Design, Automation Test in Europe
  Conference Exhibition (DATE)}}, pp. \bibinfo{pages}{1304--1309},
  \doi{10.23919/DATE51398.2021.9474232}.

\end{thebibliography}

\end{document}